 \documentclass[final,1p,times]{elsarticle}

 \usepackage{graphics}

\usepackage{amssymb}

\usepackage[flushleft]{threeparttable}

\begin{document}

\begin{frontmatter}



\title{Vortex shells in mesoscopic triangles of amorphous superconducting thin films}


\author{N. Kokubo$^1$, H. Miyahara$^1$, S. Okayasu$^2$, T. Nojima$^3$\\}

\address{$^1$Department of Engineering Science, University of Electro-Communications, Cho-fugaoka 1-5-1, Cho-fu, Tokyo 182-8585, Japan\\
$^2$Advanced Science Research Center, Japan Atomic Energy Agency,
Tokai, Ibaraki 319-1195, Japan\\$^3$Institute for Materials
Research, Tohoku University, Sendai  80-8577, Japan\\}

\begin{abstract}
Direct observation of vortex states confined in mesoscopic regular
triangle dots of amorphous MoGe thin films was made with a scanning
superconducting quantum interference device microscope. The observed
magnetic images illustrate clearly how vortices are distributed over
the triangle dots by forming not only commensurate triangular
clusters, but also unique patterns imposed by incommensurability. We
discuss the results in terms of vortex shells and construct the
packing sequence of vortices in the multiple shell structure.
\end{abstract}

\end{frontmatter}


\section{Introduction}
\label{} The confinement of a few vortices into small
superconductors provides an opportunity to study fundamental
properties of finite two-dimensional clusters in a lateral
confinement. In a circular shaped, superconducting small dot,
vortices near the edge are aligned in a ring(s) along its boundary,
while those near the disk center form a triangular cluster. These
structural ordering compete with each other and the resulted vortex
arrangement is determined by the interplay between the
geometry-induced confining potential and the mutual repulsive
interaction between vortices
\cite{Buzdin1994,BaelusPRB2002,CabralPRB2004,Misko2007}. For large
vorticities, a piece(s) of the triangular vortex lattice remains
near the disk center because of densely packed vortices
\cite{CabralPRB2004,Bolecek2014}, whereas for small vorticities,
vortices are arranged themselves into regular polygons situated on
concentric circular rings from the disk center
\cite{Buzdin1994,BaelusPRB2002,Misko2007}. This is called "vortex
shells", of which configurations have been visualized by some
experimental studies on mesoscopic superconducting disks of Nb and
amorphuous MoGe thin films \cite{GrigorievaPRL2006,Kokubo2010}.

The issue of vortex shells would be interesting when vortices are
confined in polygonal small superconductors like triangle
\cite{Zhao2008E,Cabral2009,Kokubo2015}, square
\cite{Zhao2008PRB,Kokubo2014} and pentagonal shaped dots
\cite{Huy2013}. These polygonal dots have discrete, natural
symmetries that coincide or compete with the configurational
symmetry of vortex polygons, depending on vorticity $L$. In triangle
dots, vortices intrinsically form a triangular cluster of which
symmetry is commensurate with that of the triangle geometry. This
appears when voriticity is equal to one of triangle numbers, i.e. $L
= n(n+1)/2$ with an integer $n$. For other cases, the vortex state
is frustrated with the geometry, and a metastable state(s) with free
energy close to one for the ground state is created
\cite{Zhao2008E,Cabral2009}.

Recently, we have reported the direct observation of multi-vortex
states in triangle shaped, mesoscopic dots of weak pinning amorphous
thin films with a scanning superconducting quantum interference
device (SQUID) microscope \cite{Kokubo2015}. We observed the
coexistence of multiple (metastable) patterns at some vorticities.
It turns out that the configurational degeneracy can be lifted by
introducing some deformation in the triangle geometry, i.e., the
transformation from regular triangles to isosceles ones. The
observed vortex patterns were characterized by the two-fold
reflection symmetry (incommensurate states) or the three-fold
symmetry (commensurate states), and they were interpreted as either
a commensurate triangular pattern, a linear vortex chain(s) or their
combination for $L$ up to 11.

In this study, we extend the previous work not only to study vortex
states in equilateral triangle dots for $L$ more than 11 (up to 15),
but also to understand whether vortex arrangements can be
interpreted in terms of vortex shells, as seen in circle and square
dots \cite{Kokubo2010,Kokubo2014}.

\section{Experimental}
We used a laboratory-built rf magnetron sputtering system to deposit
amorphous Mo$_{x}$Ge$_{1-x}$ films ($x \approx 75 \%$) with 0.20
$\mu$m thickness on Si substrates glued on a water cooled sample
stage. By employing conventional UV lithography and chemical/dry
etching techniques, we patterned the films into not only arrays of
triangular dots for imaging, but also long bridges for transport
measurements. The latter were used to obtain characteristic
parameters of the films as follows: The superconducting transition
temperature $T_c$ is 6.0 K, a slope of the second critical field $S$
in the vicinity of $T_c$ is $\sim$ 2.3 T/K, and the normal
resistivity at 10 K is $\sim$ 1.6 $\rm{\mu\Omega m}$. Using the
dirty-limit expression given in Ref. \cite{Kes1983}, the coherence
length at zero temperature $\xi(0)$ and the penetration depth
$\lambda(0)$ were estimated to be 5 nm and 0.5 $\mu$m, respectively.

We used a commercial scanning SQUID microscope (SQM-2000, SII
Nanotechnology) with a magnetic sensor chip integrating the Nb-based
dc-SQUIDs (Nb/Al-AlO$_x$/Nb Josephson junctions) and an inductively
coupled, small Nb-pickup loop. A homogeneous magnetic field was
applied perpendicularly to the sample surface using a coil wound in
our sample stage. The sample stage has three stepper motors which
allow us to approach and scan the samples in scales of $\mu$m with
respect to the magnetic sensor.  The vortex images presented in this
study were taken on equilibrium vortex states prepared by the
field-cool procedure, in which the magnetic field was applied in
some temperatures ($\sim$ 15 K) above $T_c$, followed by cooling the
samples in the magnetic field to temperatures of 3--4 K ($< T_c$)
for imaging \cite{Kokubo2010}.

\begin{figure}
\begin{center}
\includegraphics[width=100mm, angle=0]{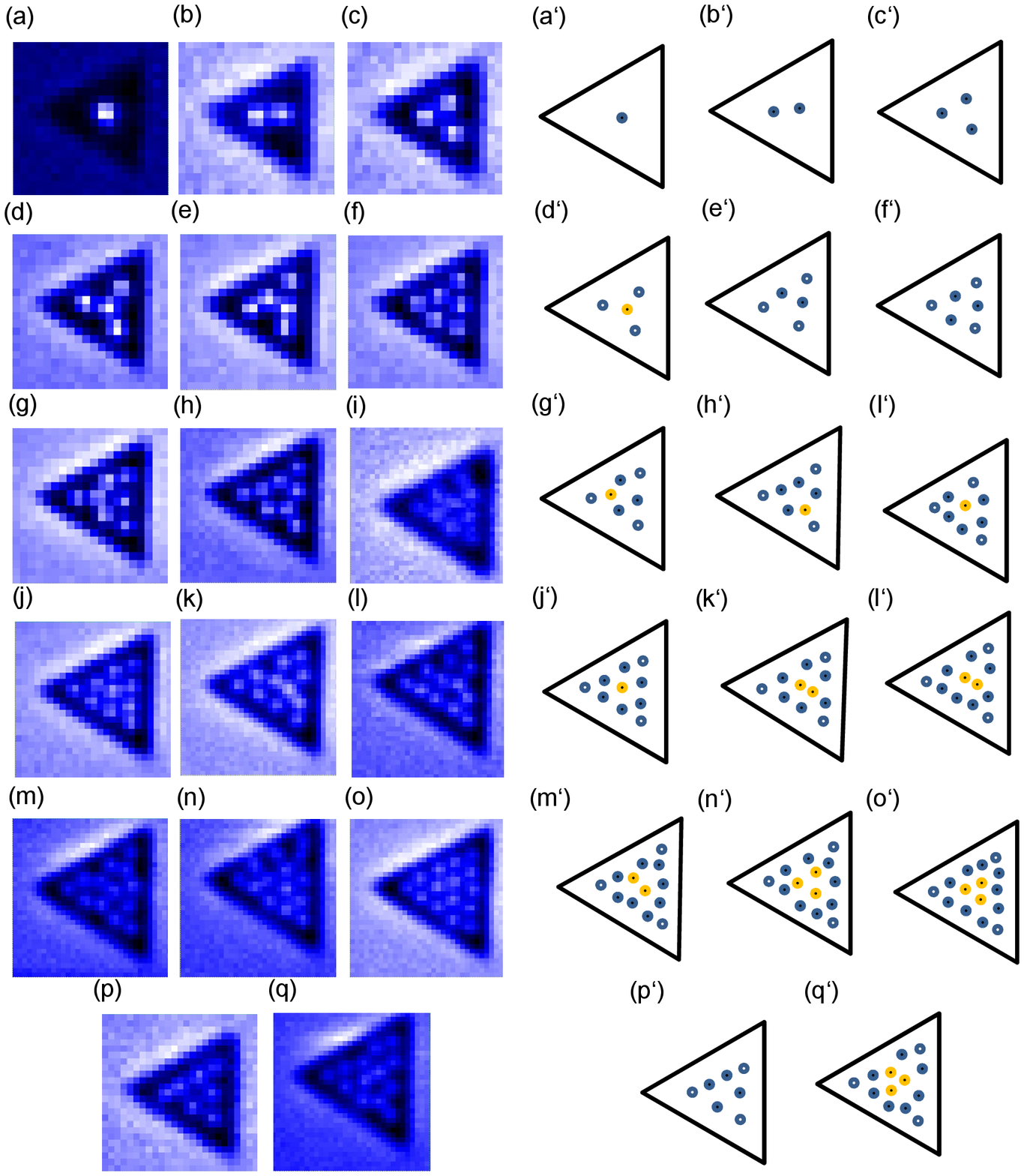}
\caption{Scanning SQUID microscopy images of vortices in 92 $\mu$m
equilateral triangular dots for vorticities $L=$ 1--15 observed at
different magnetic fields of (a) 0.50, (b) 1.50, (c) 2.50, (d) 3.00,
(e) 3.50, (f) 4.00, (g) 4.50, (h) 4.80, (i) 5.40, (j) 5.75, (k)
6.15, (l) 6.20, (m) 6.78, (n) 6.80, (o) 7.50, (p) 4.40 and (q) 6.75
$\mu$T. The traced vortex patterns are given in (a')--(q').}
\end{center}
\end{figure}

\section{Results and Discussions}
Figures 1(a)--1(o) show vortex images for vorticities $L =$ 1--15
observed in 92 $\mu$m equilateral triangle dots. These images
clearly illustrate individual vortices with reasonable spatial and
magnetic resolutions. A piece of the triangular lattice is stably
formed when the vorticity becomes the triangular number, i.e., $L =$
3 [Fig. 1(c)], 6 [Fig. 1(f)] 10 [Fig. 1(j)], and 15 [Fig. 1(o)]. As
seen partly from traced patterns in Figs. 1(a')--1(o'), vortex
arrangements are explainable in terms of the formation of triangular
vortex shells. After forming a triangle configuration at $L =$ 3,
the vortex configuration evolves with vorticity as follows; $\cdots
\rightarrow$ (1, 3) $\rightarrow$ (5) $\rightarrow$ (6)
$\rightarrow$  (1, 6) $\rightarrow$ (1, 7)$\rightarrow$ (1, 8)
$\rightarrow$ (1, 9) $\rightarrow$ (2, 9) $\rightarrow$ (2, 10)
$\rightarrow$ (2, 11) $\rightarrow$  (3, 11) $\rightarrow$ (3, 12).
Here, we employ the standard notation for multiple shells
characterized by $(N_1, N_2, \cdots, N_i)$ where $N_i$ is the
occupation of vortices in $i$-th shell from the center. Unlike
quantum atoms where new electrons are added only in the most outer
shell, both shells are filled with vortices by increasing $L$.

Many metastable states are created by the incommensurability between
the geometry and vorticity: At $L =$ 7, a single shell of (7)
appears [Figs. 1(p) and 1(p')], in addition to a double shell
configuration (1, 6) with a single vortex "core" [Figs. 1(h) and
1(h')]. At $L =$ 11 (also $L =$ 12), two configurations of double
shells were observed; one has a single vortex core, as seen in the
previous study [Figs. 2(l) and 2(x) in Ref. \cite{Kokubo2015}], and
the other has a vortex-pair core shown in Figs. 1(k) and 1(k'). At
$L =$ 13 (also $L =$ 14), double shell configurations have either
pair [Figs. 1(m) and 1(m')] or triangle cores [Figs. 1(q) and
1(q')], but the metastability disappears at the commensurate
vorticity of $L =$ 15. These vortex configurations are summarized in
Table 1.

\begin{table*}[ht]
\caption{Vortex configurations observed in mesoscopic dots with
different shapes}
 \label{Table1}
 \begin{threeparttable}
  \centering
   \begin{tabular}{cccc} \hline \hline
      $L$ &  \multicolumn{3}{c}{Configuration}  \\
    \cline{2-4}
      & Triangle & Square \cite{Kokubo2014} & Disk\cite{Kokubo2010} \\\hline
1&(1)&(1)&(1)\\
2&(2)&(2)&(2)\\
3&(3)&(3)&(3)\\
4&(1, 3)&(4)&(4)\\
5&(5)&(1, 4), (5)&(5)\\
6&(6)&(1, 5)&(1, 5)\\
7&(1, 6), (7)&(1, 6)&(1, 6)\\
8&(1, 7)&(1, 7)&(1, 7)\\
9&(1, 8)&(1, 8)&(1, 8)\\
10&(1, 9)&(2, 8)&(2, 8)\\
11&(2, 9),(1, 10)&(3, 8)&(3, 8)\\
12&(2, 10),(1, 11)&(4, 8)&(3, 9)\\
13&(2, 11),(3, 10)&(4, 9)& (3, 10)\\
14&(3, 11),(2, 12)&(4, 10)& (4, 10)\\
15&(3, 12)&(4, 11)&(5, 10) \\
16&&(4, 12)&(5, 11\tnote{a} )\\
17&&(1, 4, 12)&(1, 5, 11\tnote{a} )\\ \hline \hline
 \end{tabular}
  \begin{tablenotes}
   \item[a] A vortex(vortices) trapped at the dot edge is ignored.
  \end{tablenotes}
 \end{threeparttable}

\end{table*}

Focusing on the evolution of vortex configuration with vorticity, we
find that a single shell structure is formed not only for $L =$
1--3, but also $L =$ 5 and 6. It is worth recalling that a triangle
pattern at $L =$ 6 is commensurate with the triangle geometry and
does not accompany any observable metastable state(s). Thus, the
formation of the single shell structure after forming the double
shell structure at $L =$ 4 is robust, and therefore the packing
sequence of vortices in multiple shells is not monotonic with
vorticity.

\begin{figure}
\begin{center}
\includegraphics[width=150mm]{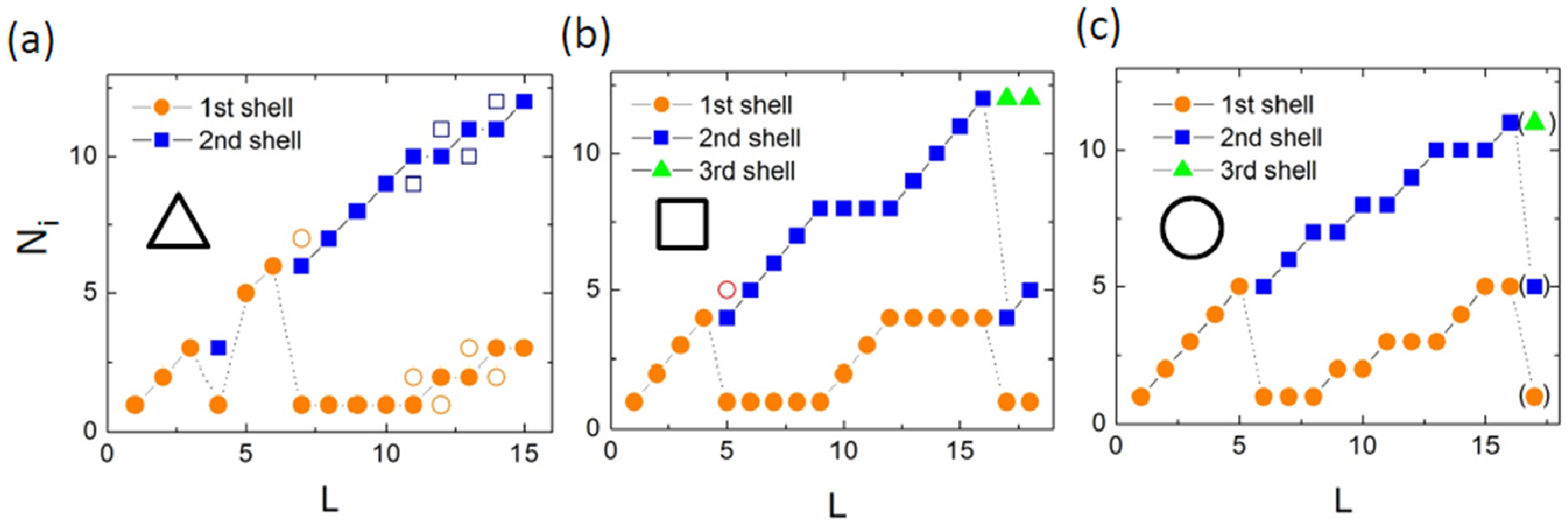}
\caption{Figure 2 (Color online) Occupation $N_i$ of different
shells vs vorticity $L$ for (a) triangle, (b) square
\cite{Kokubo2014}, and (d) circle dots \cite{Kokubo2010}. Open
symbols denote vortex configurations observed less frequently. Lines
are guide to eye.}
\end{center}
\end{figure}

This feature is also visible in Fig. 2(a), where the occupations of
vortices in different shells are plotted against vorticity. After a
monotonic increase of the occupation $N_1$ of the first shell up to
3 (at $L =$ 3), it drops to 1 due to the formation of double shell
structure at $L =$ 4. With further increasing $L$, however, the
single shell structure is reformed, and the corresponding occupation
$N_1$ jumps to 5 at $L =$ 5 and takes the maximum value of 6 at $L
=$ 6. The second shell is also discontinuous with vorticity. After
the first appearance of the second shell with $N_2 =$ 3 at $L =$ 4,
it disappears at $L =$ 5 and 6, above which $N_2$ reappears and
increases monotonically with $L$.

It is worth noting that in mesoscopic square and circle dots the
occupation of the first shell is "repeated" in their packing
sequences. As seen from Fig. 2(b), in the square dots, the
occupation $N_1$ in the first shell varies from 1 to 4 for $L =$
1--4, and this repeats for $L =$ 5--16 \cite{Kokubo2014}.  In the
circle dots [Fig. 2(c)], the variation of $N_1$ is also repeated in
a range from 1 to 5 \cite{Misko2007,Kokubo2010}. These features
indicate the presence of the maximum occupation of the first shell,
and allow us to define closed shell configurations for both dot
geometries. Namely, (4) and (4, 12) are closed configurations for
square dots, while (5) and (5, 11) are for circle dots, as listed in
Table 1. The situation is different in the triangle dots. As seen
from Fig. 2(a), $N_1$ varies from 1 to 3 for $L=$ 1--3, while it
does from 5 to 6 for $L =$ 5--6. This non-monotonic behavior of
$N_1$ seems repeatable, when one takes into account vortex
configurations for $L \geq$ 16 observed in a numerical simulation
\cite{Cabral2009}. As seen from Fig. 3 in Ref. \cite{Cabral2009},
double shell configurations with V-shaped core ($N_1 =$ 5) or
triangle one ($N_1 =$ 6) appear for $L =$ 19--22, after the
formation of a triple shell configuration ($N_1$ = 1) at $L = $ 18.
Thus, $N_1$ varies from 1 to 6 for $L =$ 7--22, but $N_1 =$ 4
remains missing.

The reason for the non-monotonic feature in mesoscopic triangle dots
originates simply from the fact that four vortices do not form a
square-like shell pattern due to the influence of the dot geometry.
Otherwise, $N_1$ would vary monotonically from 1 to 6 and the closed
shell configuration were uniquely defined as (6), which correspond
to a stable, commensurate triangle state at $L =$ 6. This may lead
to an argument if a triangle pattern with one vortex in the center
is regarded as (1, 3) or (4).




There is another filling rule of vortex shells, which is based on
the assumption that three vortices situated near the corners of the
triangle dot form the outermost shell \cite{Zhao2008E}. This model
characterizes the triangle pattern at $L =$ 6 as a double shell
configuration of (3, 3). The V-shaped pattern at $L =$ 5 becomes a
(2, 3) configuration. As a result, the corresponding packing
sequence becomes monotonic with vorticity as follows: $\cdots$
$\rightarrow$ (1, 3) $\rightarrow$ (2, 3) $\rightarrow$ (3, 3)
$\rightarrow$  (4, 3) $\rightarrow$ (5, 3) $\rightarrow$ (1, 5, 3)
$\rightarrow$ (1, 6, 3) $\rightarrow$ (1, 7, 3)  $\rightarrow$
$\cdots$.  Thus, after the formation of (1, 3) at $L =$ 4, double
shell configurations continue up to $L =$ 8, above which triple
vortex shells appear. We note that the inner parts of configurations
(ignoring three vortices in the outermost shell) are similar to ones
in disks, i.e., $\cdots \rightarrow$ (1) $\rightarrow$ (2)
$\rightarrow$ (3) $\rightarrow$ (4) $\rightarrow$ (5) $\rightarrow$
(1, 5) $\rightarrow$ (1, 6)
 $\rightarrow$ (1, 7) $\rightarrow$  $\cdots$. Thus, the observed
vortex patterns (for $L >$ 3) can be regarded as the combination of
three corner vortices and "concentric" shell configurations, but the
resultant occupations of multiple shells are not repeatable due to
the limited occupation of the outermost shell.

\section{Summary}
We have presented SQUID images of vortex states confined in
mesoscopic regular triangle dots of weak pinning amorphous MoGe
films for vorticities $L =$ 1--15. The formation of a triangular
vortex cluster occurs at the triangle numbers of vorticities ($L =$
3, 6, 10 and 15), while for other vorticities, vortex patterns are
mostly frustrated with the dot geometry, and form unique
arrangements, including metastable ones, determined by the
incommensurability between the geometry and vorticity. Irrespective
of the (in)commensurability, the observed vortex arrangements are
explainable in terms of vortex shells. The corresponding packing
sequence reveals that the single shell structure is stably reformed
after the formation of the double shell structure. Thus, closed
shell configurations are not uniquely defined, in contrast to the
results in mesoscopic circle and square dots
\cite{Kokubo2010,Kokubo2014}. We believe that the observed feature
is not related to the metastability, but intrinsic to mesoscopic
triangle dots.
\bigskip

\noindent{\bf Acknowledgment} This work was supported by JSPS
KAKENHI Grant Numbers 23540416, 26287075 and 26600011,
Nanotechnology Network Project of the Ministry of Education,
Culture, Sports, Science and Technology (MEXT), and the
Inter-university Cooperative Research Program of the Institute for
Materials Research, Tohoku University (Proposal No. 14K0004).



\end{document}